# FORUM ON IMMUNE DIGITAL TWINS: A MEETING REPORT

Reinhard Laubenbacher[1*], Fred Adler[2], Gary An[3], Filippo Castiglione[4], Stephen Eubank[5], Luis L. Fonseca[1], James Glazier[6], Tomas Helikar[7], Marti Jett-Tilton[8], Denise Kirschner[9], Paul Macklin[6], Borna Mehrad[1], Beth Moore[9], Virginia Pasour[10], Ilya Shmulevich[11], Amber Smith[12], Isabel Voigt[13], Thomas E. Yankeelov[14], Tjalf Ziemssen[13]

[1]Department of Medicine, University of Florida, Gainesville, FL
[2]Department of Mathematics and School of Biological Sciences, University of Utah, Salt Lake City, UT
[3]Department of Surgery, University of Vermont, Burlington, VT
[4]Biotechnology Research Center, Technology Innovation Institute, Abu Dhabi, United Arab Emirates
[5]Biocomplexity Institute and Initiative, University of Virginia, Charlottesville, VA
[6]Department of Intelligent Systems Engineering, Indiana University, Bloomington, IN
[7]Department of Biochemistry, University of Nebraska, Lincoln, NE
[8]U.S. Walter Reed Army Institute of Research, Silver Spring, MD
[9]Department of Microbiology and Immunology, University of Michigan, Ann Arbor, MI
[10]U.S. Army Research Office, Research Triangle Park, NC
[11]Institute for Systems Biology, Seattle, WA
[12]Department of Pediatrics, University of Tennessee Health Science Center, Memphis, TN
[13]Center for Clinical Neuroscience, Carl Gustav Carus University Hospital, Dresden, Germany
[14]Department of Biomedical Engineering, Oden Institute for Computational Engineering and Sciences, Departments of Biomedical Engineering, Diagnostic Medicine, Oncology, The University of Texas, Austin, TX, and Department of Imaging Physics, The University of Texas MD Anderson Cancer Center

[*]Corresponding author, reinhard.laubenbacher@medicine.ufl.edu

**ABSTRACT**
Medical digital twins are computational models of human biology relevant to a given medical condition, which can be tailored to an individual patient, thereby predicting the course of disease and individualized treatments, an important goal of personalized medicine. The immune system, which has a central role in many diseases, is highly heterogeneous between individuals, and thus poses a major challenge for this technology. If medical digital twins are to faithfully capture the characteristics of a patient's immune system, we need to answer many questions, such as: What do we need to know about the immune system to build mathematical models that reflect features of an individual? What data do we need to collect across the different scales of immune system action? What are the right modeling paradigms to properly capture immune system complexity? In February 2023, an international group of experts convened in Lake Nona, FL for two days to discuss these and other questions related to digital twins of the immune system. The group consisted of clinicians, immunologists, biologists, and mathematical modelers, representative of the interdisciplinary nature of medical digital twin development. A video recording of the entire event is available. This paper presents a synopsis of the discussions, brief descriptions of ongoing digital twin projects at different stages of progress. It also proposes a 5-year action plan for further developing this technology. The main recommendations are to identify and pursue a small number of promising use cases, to develop stimulation-specific assays of immune function in a clinical setting, and to develop a database of existing computational immune models, as well as advanced modeling technology and infrastructure.

**INTRODUCTION**

The concept of a *medical digital twin* (MDT) represents a pivotal technology envisioned to make personalized medicine a reality. This entails using predictive computational models to harness diverse patient data over time, allowing for identification of optimal interventions and corresponding predictions of their effectiveness for an individual patient; see, e.g., [1][2][3][4]. Scaling up this concept into a widely used medical technology necessitates substantial coordinated advancements across several fields, including human biology, medicine, biochemistry, bioinformatics, and mathematical and computational modeling. A sign of increasing interest in this technology was evident in the workshop "Opportunities and Challenges for Digital Twins in Medicine," organized by the National Academies of Science, Engineering, and Medicine in January 2023 [5][6]. One possible long-term vision is a virtual replica of an entire patient that evolves with the patient over the course of their lives, as articulated by the Virtual Physiological Human Institute [7] and the European Virtual Human Twin Project [8]. The foundations for MDT technology, however, are yet to be developed. The Forum described here, and other efforts [9][10] have focused on digital twins for medical conditions related to the immune system. This provides a narrower focus, but at the same time addresses a wide range of diseases that involve the immune system in an essential way, such as infectious diseases, autoimmune diseases, and cancer, among others.

For this article, we adopted a broad definition of an MDT: it comprises a patient, a set of temporal data collected from that patient, and a computational model calibrated with this data. This allows for conclusions to be drawn about the patient, either at one time point or in the form of outcome forecasting, with or without interventions. In essence, the patient is paired with a computational model that is personalized and may synchronize repeatedly with the patient over time. Alternatively, the MDT might represent a patient population for the purpose of virtual clinical trials. This ongoing linkage and exchange of data and information between the patient and the MDT is the most important characteristic that distinguishes an MDT from a model, even a personalized model. This concept applies to treatment of an existing health condition or preventing the emergence of one, in which case the data might come from electronic health records or wearable sensors [11]. This definition of personalization is rooted in the hypothesis that any two given patients will differ in underlying biology, disease trajectory, and hence optimal treatments over time. Even with this broader definition, there are relatively few instances of MDTs that have reached patient care. While there are many models in the literature that could be further developed into MDTs, the link to individual patients has not been fully explored in most cases, so that extensive further development is required. The Forum, the subject of this report, was primarily focused on such early stage MDTs and what is needed to progress to the clinical stage. The report highlights some projects of this kind as use cases: **computational models that are being used or could be further developed for use in informing the treatment of individual patients.**

While the industrial application of the digital twin concept is instructive, it differs from medical digital twins in several key aspects. Most importantly, human biology is not the result of a planned design, but the outcome of an evolutionary process, with many emergent properties. We do not have a complete theoretical understanding of biological systems, providing a list of general principles that could form the basis of computational models, as we do for physical systems. Finally, two other characteristic features of biological systems are genotypic and phenotypic heterogeneity across individuals and stochasticity in system dynamics. All these features present massive challenges to mathematical modeling of individuals and populations. A wide range of mechanistic, phenomenological, and statistical models are being used for this purpose. Biological mechanisms cross scales as do therapeutic interventions. For instance, many drugs target intracellular mechanisms but have tissue- or organ-level effects. Therefore, many mechanistic MDTs will need to span multiple scales. This raises the question of whether our current repertoire of modeling paradigms is sufficient to form the basis of digital twins across various health conditions and how to choose the right type of model for each one. How can we effectively and credibly capture key features of human biology in a manner suitable for a specific clinical application while considering the diversity of patients and their individual characteristics?

To begin addressing these questions, an international group of experts convened in Lake Nona, FL, February 23-24, 2023, for the "Forum on Precision Immunology: Immune Digital Twins" [12], supported by a grant from the Biomathematics Program at the U.S. Army Research Office. The aim was to discuss these questions and assess examples of ongoing modeling projects that are part of MDT development related to immunity. This report encapsulates a synthesis of these discussions, offering a sample of ongoing MDT projects at different stages of development, and an outline of challenges to be addressed over the next five years.

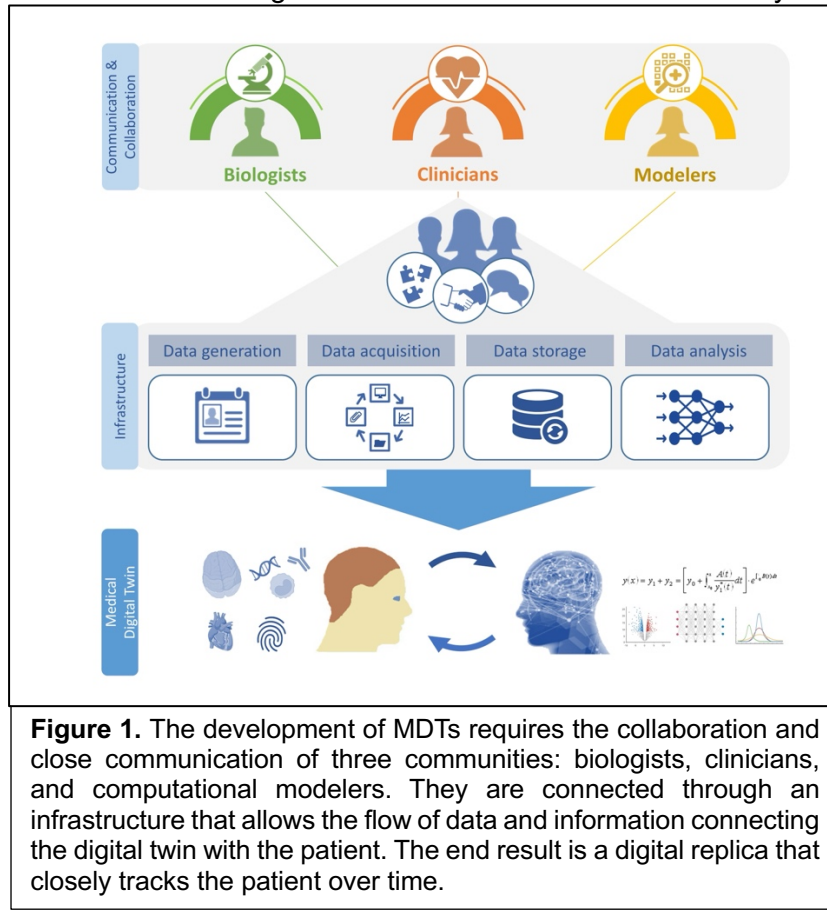

**Figure 1.** The development of MDTs requires the collaboration and close communication of three communities: biologists, clinicians, and computational modelers. They are connected through an infrastructure that allows the flow of data and information connecting the digital twin with the patient. The end result is a digital replica that closely tracks the patient over time.

The development of MDTs takes place at the interface of medicine, experimental biology, and mathematical modeling (see Fig. 1). The Forum participants are all authors of this article, and represent a cross-section of these fields, including clinicians, immunologists, experimental biologists, and mathematical modelers. The Forum served as a venue to discuss the different perspectives each of these communities has on the prospect of using personalized computational models in the clinic. To facilitate an exchange of ideas across these fields, the program consisted of a collection of 45-minute blocks, with a 15-minute presentation by a participant, followed by 30 minutes of discussion. The only audio-visual aid available to presenters was a whiteboard, favoring discussion over formal presentations. High-quality audio-visual recordings of the individual sessions are available through links at [12]. The reader is encouraged to view the presentations, as they contain many valuable ideas, viewpoints, and information not contained in this synopsis. Below are the titles of each of the discussion sessions. (The titles correspond to the links to video recordings on the website [12].)

**Adler:** Summary of conceptual, scientific, practical and ethical challenges and opportunities discussed by other participants in developing medical digital twins.
**An:** Axioms of personalized precision medicine.
**Castiglione:** Constructing a computational representation of the Immune System: necessities, constituents, and operational aspects, along with proposed approaches for model development.
**Eubank:** Lessons to be learned from other fields about data assimilation.

**Glazier:** A theoretical framework for the construction of medical digital twins.
**Helikar:** Towards a General Purpose Immune Digital Twin.
**Jett-Tilton:** Digital twins for PTSD.
**Kirschner:** Models and Tools for building beta versions of digital partners.
**Laubenbacher:** Introduction to the Forum.
**Macklin:** Integration of standardized, reusable descriptions of cell behaviors and interactions.
**Mehrad:** The application of MDTs to the intensive care unit.
**Moore:** Immunologic considerations for building MDT.
**Pasour:** Funding opportunities.
**Shmulevich:** Patient Digital Twin for Acute Myeloid Leukemia.
**Smith:** Immune heterogeneity in the context of lung infection.
**Yankeelov:** Imaging-based digital twins for oncology.
**Ziemssen:** A digital twin for autoimmune diseases accessible to the patient.

We now outline the general themes of the Forum discussions for each of the three pillars of MDTs: the clinic, immunology, and mathematical modeling. And we extract a collection of action items for a 5-year plan to further MDT development.

**HUMAN IMMUNE SYSTEM BIOLOGY**
The human immune system is highly specialized and has evolved to have exquisite specificity for defending its host from injury and infection. During health, the immune response is tightly orchestrated to respond to threats without inducing significant tissue damage, but dysregulation can occur, contributing to cancer or autoimmunity [13]. The complexity, specificity, and regulation are all challenges to creating effective MDTs involving the immune response. A few of these challenges and opportunities are highlighted below. The list is not comprehensive and serves to illustrate some key directions for research and data collection.

1. Genetic Diversity and Immune Cell Activation: Self-specificity of the adaptive immune response in any individual is governed by the ability of the host cell to display peptides derived from foreign threats (*e.g.*, microbes) in the context of human leukocyte antigens (HLA). Every person inherits 6 major HLA alleles from each parent (HLA- A, B, C, DP, DQ and DR for 12 total) and there are over 37,000 HLA and related alleles characterized to date [14]. If this was relevant information for a particular MDT, then we could simply tissue-type individuals and feed their alleles into our MDT. However, while there are good computational tools to predict what peptide will fit into the groove of the appropriate HLA molecule of each person [15], we do not yet have good methods to predict which of the many possibilities is likely to be immunodominant within a person. The immunodominance of the response will be related to the cadre of T and B cell receptors that are present in each individual. It is estimated that $10^{13}$-$10^{18}$ different T and B cell immune receptors are generated through genetic rearrangements, reassortments and editing; however, many of the cells carrying autoreactive receptors (or those that don't recognize self at all) are deleted, leaving approximately $10^{11}$-$10^{12}$ different specificities in circulation [16]. The ability of a particular T cell to encounter a particular antigen-presenting cell with the correct HLA and the correct peptide for activation is stochastic in nature and involves the probability of the two cells encountering one another and the strength of the interactions and co-receptor signaling to activate the cells. Once activated, a further level of complexity involves the cytokine stimulations that will direct the T cells into particular subsets with unique attributes [17]. Thus, the sheer complexity in terms of genetics, random interactions, cytokine profiles, receptor diversity and outcomes are daunting when considering how to model individualized immune responses within an MDT.
2. Phenotype ≠ Function: Despite the immense complexity described above, there are emerging machine learning and artificial intelligence algorithms that can predict peptide binding to HLA, T cell binding to peptide-HLA and T helper cell differentiation programs [18][19][16][20][21], but it may not be necessary to build all the inherent complexity into an MDT. For example, RNA sequencing technology and multi-parameter flow cytometry have given us the opportunity to finely phenotype immune cell subsets (with flow cytometry being a potentially real-time source of data), but the molecular phenotype of the cells does not necessarily connote function. Thus, an area to focus on for the future is to develop rapid functional assays that will assess a desired output (*e.g.*, production of interferon gamma in response to a viral antigen stimulation), which would indicate that antigen-specific T cell responses had occurred. Identification of one or a few critical functional assays of relevance to the disease process being modeled may be an effective way to aggregate the many variables of immune activation into a single continuous variable for modeling immune response.
3. Stochasticity: In medicine, when patients respond differently to various infections or treatments, we often attribute this to differences in age, genetics, or other lifestyle factors. However, even in the research laboratory where experiments can be conducted in genetically identical individuals (*e.g.*, mice), housed in identical conditions and infected with identical pathogens and doses, we see variations in the response. For example, for any infection, researchers can generate an $LD_{50}$ which is defined experimentally as the dose at which 50% of the animals die. How is it possible for there to be a dichotomy in the response between identical individuals? This is generally attributed to stochastic events such as deposition of the infectious agent in particular regions of the body (*e.g.*, different lung lobes), and it may relate to the number of immune cells patrolling that particular area of deposition. Given that this is a known biological phenomenon, such stochasticity will need to be built into the MDT.
4. Source of samples: Another hurdle for MDTs is that the most easily accessible source of biological material is from the bloodstream; however, the response patterns in circulation often do not mirror the tissue-specific responses. Finding ways to safely sample tissues like the brain or lung or heart remain a challenge, and therefore surrogate measures are likely needed. MDTs can be built from electronic health records, but these will be incomplete for many tissue compartment responses.

**Five-year action plan for human immune system biology**
A major goal for the next 5 years should be the development of enhanced *in vitro* and *ex vivo* assays to accurately predict immune function in response to relevant stimulation. Additionally, creating workflows that can swiftly provide this crucial information in a clinical setting will aid in prognostication in the MDT. Achieving this will likely necessitate improved modeling of *in vivo* conditions, including aspects such as oxygen saturation, tissue architecture and tissue-specific compartmentalization of responses. One potential approach could involve monitoring an individual's "immune baseline" through routine blood work at homeostasis. This may sufficiently inform the development of MDTs. While this could help predict some clinically relevant responses (*e.g.*, someone with high basal IL-4/IL-13 levels and high circulating eosinophil levels might be predicted to have more exuberant allergic responses), the post-stimulation response beyond the baseline may be more pertinent. Therefore, we strongly advocate for the development of stimulation-specific assays of immune functions.

Other recommendations for improved modeling of the human immune response via MDTs include:
- Development of *ex vivo* culture systems that rapidly measure relevant cell behaviors and interactions (proliferation, cytokine secretion, phagocytosis, etc.) in response to disease-relevant stimuli to provide data for MDT modeling.

- Identification of culture systems and animal models that more closely mimic human *in vivo* biology (*e.g.*, 3-D cultures with multiple cell types, organoids, hypoxic environments) to provide more relevant insight for modeling. The inclusion of immune cells in *ex vivo* tissue models, which is currently rare, should be encouraged.
- Identification of ways to sample tissue-specific compartment responses (*e.g.*, using implanted sensors or scaffolds for sampling). Alternatively, identification of surrogate markers for tissue-specific responses in circulation.

**THE CLINIC**
The "twin" component of an MDT explicitly ties the digital object to an individual patient, and therefore inherently incorporates a translational purpose of the MDT. As such, the potential clinical role of an MDT will drive its development. Clinical practice can be divided into a series of distinct, but related tasks: 1) diagnosis of a potential disease state (this includes monitoring a state of health to identify divergences); 2) prognosis, which attempts to predict or forecast a particular disease trajectory; 3) personalization/optimization of existing therapies; and 4) the discovery/testing of novel treatments. Items 1-3 form the basis of current clinical practice, with a mixture of basic pathophysiology, evidence-based (ideally) practice guidelines and an individual physician's expertise and intuition. Conversely, Item 4, the discovery/testing of novel treatments, is traditionally the purview of research. These tasks can also be grouped into types: 1) a classification task ("What illness is the medical team dealing with?"); 2) a forecasting task ("What is going to happen to my patient in the future?"); and 3) a control task ("What is the best course of action to make my patient better?"). Classifying a particular use-case for potential MDTs can aid in determining what sort of data is necessary and available (or not) for a particular purpose, what the time scale might be for the updating between the MDT and the real-world twin, and what type of computational method(s) would be needed to propagate the MDT forward in time (this aspect will be covered in more detail in the "Mathematical and Computational Modeling" section below). Another application one could envision is for an MDT to serve as a benchmarking tool to evaluate current therapy.

It is worth stating explicitly that MDT technology will likely follow the same path as other new technologies. Initial prototypes will have a limited range of capabilities and modest performance and serve perhaps more as a proof-of-concept than fully functional products. The minimum bar any MDT will have to clear, of course, is that it needs to perform at least as well as the standard of care for a given application, without any additional risks to patients. The experience gained from initial development and data collected from its use will then drive the development of increasingly more sophisticated versions.

As an example, we present a cascading set of increasingly powerful potential use cases of MDTs in the treatment of sepsis, one of the largest sources of morbidity, mortality and health care costs world-wide (WHO).
1. <u>Early detection of sepsis is a health-monitoring, classification task</u>. This could employ an MDT trained on physiological signals, electronic medical record data and standard laboratory values to deliver an "early warning system" for sepsis.
2. <u>Predicting the trajectory of sepsis</u>. This could be related to the diagnosis task, as certain features might suggest a clinical trajectory that leads to sepsis. It could also be applied to patients already diagnosed with sepsis, to attempt to risk-stratify patients to identify those at risk for clinical deterioration.
3. <u>Optimization of existing therapies for sepsis</u>. The mainstay of current treatment of sepsis involves early administration of antibiotics, source control of potential sources of infection, and physiological support, which includes fluid resuscitation, the use of vasopressors to support blood pressure, and mechanical devices to support failing organs (*i.e.*, ventilators and dialysis machines). The combinations of applications, both in time and in degree, could be guided by a sufficiently trained MDT.

4. <u>Discovery and deployment of new therapies</u>. The unfortunate fact of sepsis is that, to date, there is no generally accepted means of interrupting the underlying inflammatory/immune biology that drives sepsis and its subsequent organ failure. Major contributing reasons for this are the overall heterogeneity of the septic population (reflected in a gap between the means of "diagnosing" sepsis and the degree of knowledge regarding the cellular-molecular mechanisms that drive the disease) and the complexity, both in terms of the underlying biological mechanisms and their dynamics in given different insults, of the disease course. In short, effective treatment/control requires identifying the right patient at the right time for the right set of therapies, and the current means of doing these tasks for a septic patient are woefully inadequate. It is here that MDTs can play an invaluable role in personalizing the characterization of a septic patient so that "right patient, right time, right drug(s)" can be achieved.

**Five-year action plan for the clinic**
A Five-Year plan for the development and deployment of MDTs needs to integrate capabilities that can improve patient health within a decade, with aspirational capabilities that will allow MDTs to reach their full potential. With this in mind, we propose the following actions:
1. Clear identification of specific disease processes to be targeted for development (some candidates are identified below).
2. Explicit definition of specific use-cases/tasks for a given disease.
3. Identification of specific data types required for each use case, whether that data currently exists in some form or will be available in the future to meet the capabilities of an aspirational MDT. Of note, obtaining time series/ongoing collected data is essential to this step, as the concept of time-evolution of the MDT is inherent to its definition.
4. These first three steps should be integrated into a detailed "roadmap" for the development and deployment of the MDT.
5. Use of this roadmap to engage collaborators and stakeholders (*i.e.*, clinicians and clinical researchers, assay developers, mathematical modelers, and biologists) to facilitate the collection of existing data and develop the capability to acquire new types of data as needed.
6. Deployment of an initial MDT with diagnostic and prognostic utility for clinical decision-support. Ideally, there should be enough preliminary data such that reasonable planning could be implemented after a short period for clinical trials to demonstrate their utility.

**MATHEMATICAL AND COMPUTATIONAL MODELING**
The engine of any MDT is a computational model. Depending on the application and available data, it may include mechanistic information about the relevant human biology, and it may take as input information specific to either an individual patient or a patient population. In all cases, the output is information that can be used in the treatment of an individual patient. Figure 2 depicts the role of the computational model in the workflow of MDT applications.

For instance, a deep learning model might be trained on clinical data from a large patient cohort of gastric cancer patients, and is then used to determine a patient's response to an immunotherapy treatment [22]. Such models may or may not include any mechanistic information about the relevant tumor biology, such as mutated signaling pathways and their downstream effects, and predictions for the specific patient are based on correlation between the patient's data and those of the reference population used in the model. At the other end of the spectrum, a computational model may capture all known features of human biology relevant to a given application and may make treatment recommendations based on a model analysis, informing clinical trials, without using any data from a specific patient [22]. An MDT that most closely adheres to the industrial concept of a digital twin needs to do both. It will use a mechanistic model of some aspect of human biology and also provide individual treatment recommendations based on that patient's data. **The focus of Forum participants was primarily on MDTs based on a mechanistic computational model.** This preference stems from the ability of mechanistic MDTs to link outcomes to mechanisms, thereby informing treatment. Additionally, these models allow for the performance of uncertainty quantification in relation to their predictions.

Many mechanistic models of human biology are now available, particularly those incorporating aspects of the immune system. For numerous applications, the underlying model of an MDT will need to encompass various mechanisms, spanning several spatial and temporal scales. For example, while most drug mechanisms are intracellular, their effects manifest at the tissue or organ scale, necessitating cross-scale integration. The immune response to an infection is multifaceted, coordinating diverse mechanisms and cell types. Consequently, computational models for MDTs will likely be high-dimensional, multi-scale, multi-physics, hybrid, and stochastic, containing numerous parameters. Integrating heterogeneous data types, from molecular to physiological, will be essential for their parameterization and application. Most crucially, these models should be adaptable to individual patient data. Very few such models have been constructed for clinical use or new biology discovery, leading us into uncharted territory in their construction, analysis, validation, and application. Below is a proposed 5-year plan to develop the necessary technology for building credible MDTs for applications involving the immune system.

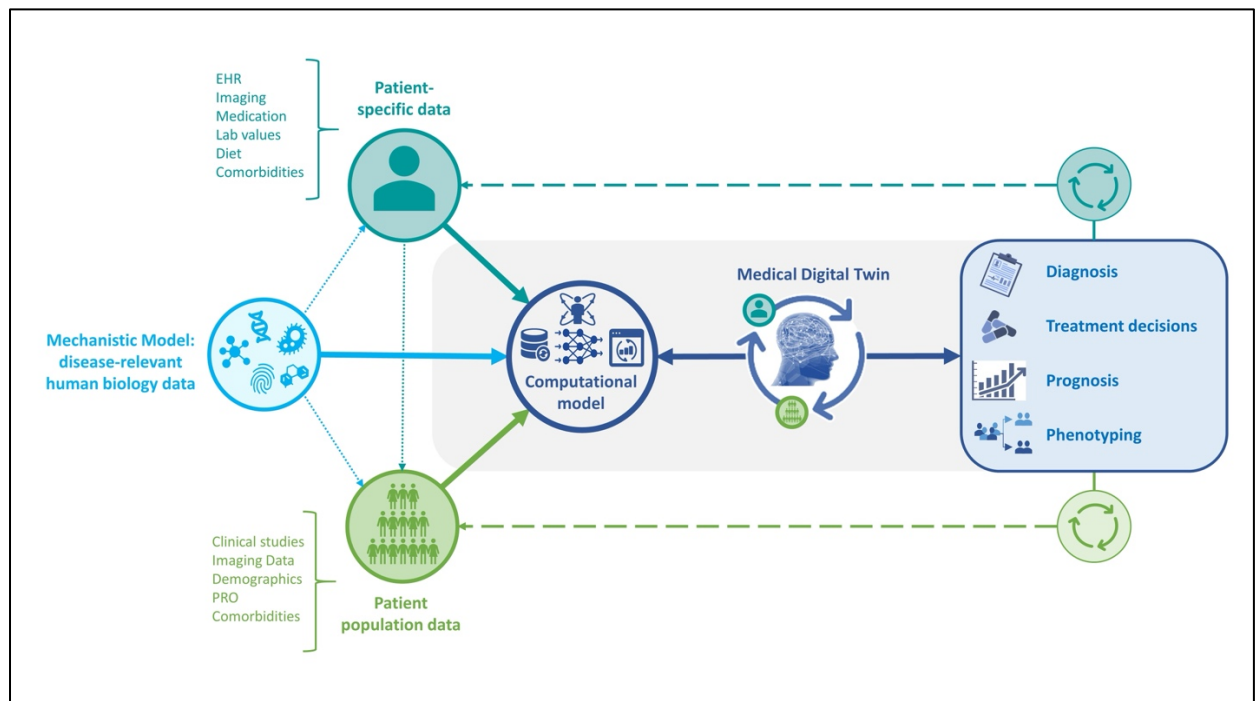

**Figure 2.** The computational model at the heart of an MDT serves several purposes. It integrates human biology, clinical data, data characterizing reference populations, and patient-specific data. It is personalized to the patient and is periodically re-calibrated. Control algorithms attached to the Model can be used to optimize available patient treatments.

**Five-year action plan for mathematical and computational modeling.**
1. The biomedical modeling community has spent decades building complex models of different medical and disease processes in humans from cancer to infections. These are all potentially usable as drivers of MDTs or components thereof. As a first step, we need to develop and curate a repository of model templates (*i.e*., accepted model structures) and specific model modules (*e.g.*, peer-reviewed models of specific signaling networks) that can be used in the construction of MDTs, ranging from intracellular to physiological scales. Existing repositories include, e.g., Biomodels [23], Cell Collective [24] and GinSim [25]. These can be built upon for a more comprehensive curated collection.
2. The most important criterion for models underlying MDTs is their credibility. Much effort has gone into developing criteria and rules for model credibility [26][27][28]. For models that are used in the treatment of patients, the set of rules will need to be modified. The final judgment whether a model is credible lies with the clinician who uses the MDT as a decision support tool. This requires a higher standard than for models used to discover new biology or even for models used in drug development, where the clinical trial is the final arbiter.
3. Existing techniques for the validation, calibration, and analysis of computational models, most importantly sensitivity and identifiability of model parameters, are not always directly applicable to stochastic multiscale hybrid models or can be computationally expensive to apply. Research is needed to develop appropriate model analysis techniques for MDTs.
4. For many applications, MDTs will be used to forecast the future health trajectory of a patient, as well as the effect of available interventions to change it. Existing approaches to forecasting and data assimilation, such as methods based on Kalman filters used in numerical weather prediction, have several limitations when applied to high-dimensional hybrid models. Existing control and optimization methods (*e.g.,* [29]) mostly apply only to ordinary differential equations models. Research is needed to develop novel forecasting and control approaches suitable for complex MDTs.
5. There are many existing models of disease processes and immune system function that can be used to build MDTs, as mentioned above. Research is needed to develop a platform for the modular construction of complex MDT models from component models. Such a platform is essential for achieving the long-term vision of a virtual patient. A possible approach includes [30].
6. For applications, ensuring that MDT simulations are conducted within a clinically relevant time frame is crucial. This often necessitates the use of high-performance computing resources. Additionally, it may require the development of approximate models that, while offering rapid simulation capabilities, still maintain a high level of accuracy.
7. If MDTs are to be used in patient care, they will need to be accessible by clinicians and patients through appropriate user interfaces. The user will also need to be able to evaluate the trustworthiness of MDT forecasts and recommendations.

**EXAMPLES OF ONGOING MDT PROJECTS**
The wide range of MDT projects and application areas would require a comprehensive review of the subject. Here, we present a selection of ongoing projects by some of the Forum participants. The selection illustrates several disparate types of applications, methodologies, and uses. They are at different stages of development and collectively illustrate the issues we have raised in this meeting report/perspective. A summary of the projects presented at the Forum can be seen in Table 1. Generally, projects can be characterized as follows:
1. Whether the underlying computational model/specification is generated using an existing modeling toolkit/format (which would allow for potentially greater community level expansion) or a "custom" model specific to a particular research laboratory.
2. The disease process addressed by the nascent MDT project.
3. The data types and sources that are available for the data interface between the patient and the digital twin. This ranges from demographic and clinical descriptive data, as found in electronic medical records, the results of diagnostic imaging and tests, and more specific assays that are currently mostly available in the research context (*i.e.*, gene expression, multiplexed mediator assays or highly granular cell type characterization).
4. Whether such a data interface currently exists for the nascent MDT at its current level of development.
5. The modeling method used for the current computational model/specification of the MDT. This includes whether a mechanism-based dynamic model is used, whether a machine learning/artificial intelligence component is part of that dynamic model, or whether the specification is in its early development stages.
6. The approach by which the MDT computational model/specification is personalized (*i.e.*, the "twinning" process) to an individual patient in the real world. A precursor to the actual personalization would be the generation of virtual populations, which represents a theoretical distribution of real-world individuals, but have not yet reached a point of development where there can be a direct mapping/connection to an individual patient in the real world.
7. The ostensible clinical goal of the MDT. This could range from diagnosis/surveillance, prognosis/disease trajectory forecasting, optimization and personalization of existing therapies, or the discovery of novel therapies, be they new therapeutic agents, new combinations of existing drugs, or the repurposing of existing drugs into new disease contexts.
8. Whether the MDT project has a patient-facing/engaging interface. This step informs whether, based on the context of its use, such a patient-engagement capability would increase the willingness of potential patients to participate in the MDT project, and helps establish a context for dealing with ethical issues such as patient privacy, data ownership/stewardship and participatory medical decision-making.

**A host model for tuberculosis that spans the molecular to the whole host scale** (D. Kirschner)

Tuberculosis (TB) continues to be a global disease threat, even compared to the COVID pandemic. Approximately one-fourth of the world is infected with *Mycobacterium tuberculosis* (Mtb) in their lungs; however, most patients are classified as having latent tuberculosis (~90%) with only a small percentage with clinically active disease (~10%) (WHO). While patients are categorized within what seems as binary states, recent work has shown that TB manifests as a spectrum of outcomes within both humans and non-human primates (NHPs) [31][32][33][34]. Importantly, latently-infected individuals may undergo reactivation events and thus serve as a potential reservoir for transmission [35][36]. Much remains unknown about the biology that drives disease states in pulmonary TB. Understanding what drives different infection outcomes is important as it will inform development and approaches for treatment and prevention.

      The hallmark of TB is the formation of lung granulomas, which are organized immune structures that immunologically constrain and physically contain Mtb. These develop in the lungs of infected hosts after inhalation of mycobacteria [31]. NHP data have shown that a single mycobacterium is sufficient to begin the formation of a granuloma and that each granuloma has a unique trajectory. Granulomas are composed of bacteria and various immune cells, such as macrophages and T cells (primarily CD4+ and CD8+ T cells, although other unconventional T cell phenotypes are also present). T cells have well-known critical functions against Mtb [37], but unlike other infections, T cells are slow to be recruited to the lungs, arriving approximately a month after infection. Lung-draining lymph nodes (LN) serve as the sites for initiating and generating an adaptive immune response against most infections, including Mtb.

      We developed a novel whole-host scale modeling framework that captures key elements of the immune response to Mtb within three physiological compartments - LNs, blood and lungs of infected individuals. Together, this model platform, called *HostSim* [38][39], represents a

whole-host framework for tracking Mtb infection dynamics within a single host across multiple length scales and long time scales (days to months to years). We calibrated and validated the model using multiple datasets from published NHP studies and humans. *HostSim* offers a computational tool that can be used in concert with experimental approaches to understand and predict events about various aspects of TB disease and therapeutics.

Recently, we have generated hundreds to thousands of *HostSim* "virtual patients" that are infected with TB at different times and have slightly unique immune characteristics. We refer to this collection of virtual hosts as a "virtual cohort". This virtual cohort can serve as a bank of digital "partners" that can be closely associated with an actual patient. Initially, a large group of partners (*i.e.*, a 'digital family') would be assigned to that patient. Then, as more data become available, the family of partners that are associated with this patient would narrow until a single digital twin remains.

**Virtual Patient Cohorts for Virus Infections** (A.M. Smith)
Respiratory viruses cause a significant number of illnesses and deaths each year, with considerable health and economic burden. Infections with viruses like influenza or SARS-CoV-2 yield a variety of outcomes that range from asymptomatic to fatal. Numerous viral and host factors in addition to complications from other pathogens and underlying diseases can result in heterogeneity in the severity of infection, but their contribution or those from other, hidden mechanisms is unknown. This makes predicting a patient's disease trajectory and the potential for efficacious vaccination or antiviral therapy challenging. The goal of this project is to build virtual patient cohorts (VPCs), with each patient having a personalized immune trajectory [40] to define immunologic processes that initiate diverse outcomes. We construct mechanistic and experimentally validated computational models of the host response and define immune correlates of disease. A focus is on establishing the nonlinearities that drive many immune processes and their connections to disease [41][42]. Within this approach, models are iteratively updated with new data, as immunological knowledge evolves, and as smaller models are validated with targeted experimentation alongside generating diverse VPCs to evaluate underlying comorbidities.

**The Digital Twin Innovation Hub** (T. Helikar)
The Digital Twin Innovation Hub [43], established in August, 2022, is leading the development of a general purpose immune digital twin that will be contextualizable and applicable to many, and eventually any, immune-related pathology. A comprehensive cellular-level model and map of the immune system, consisting of nearly 30 cell types, over 30 cytokines and immunoglobulins spanning both innate and adaptive immunity has been developed to form a "blueprint" of the general purpose immune digital twin [44]. Detailed sub-cellular models of signal transduction and genome-scale metabolism for each of the 30 cell types have also been developed (e.g., dendritic cells, CD4+ T cells [45][46]. Work to integrate these sub-cellular models into a comprehensive multi-scale, multicellular model of the immune system is under way.

Digital Twin Innovation Hub is also developing a software infrastructure to enable the construction, contextualization, personalization, analysis, and simulation of the general purpose immune digital twin. To accomplish this, the Hub is leveraging and building atop of Cell Collective, a web-based collaborative modeling platform [24]. To this end, Cell Collective supports several modeling approaches, including logical, kinetic, and constraint-based models, and will soon also support physiologically-based pharmacokinetic/pharmacodynamic models and virtual clinical trials. Cell Collective also provides a repository of computational models, which will provide a gateway to features that will enable their integration into multiscale systems - medical digital twins.

A key principle of Cell Collective is its broad accessibility. To fully leverage the potential of medical digital twins, it will be critical that the technology is accessible to a wide range of user audiences, including translational researchers, clinicians, and patients. As such, in Cell Collective, no mathematical or programming skills are required for users to build, modify, simulate, or analyze models. It also allows users to focus on the mechanistic information used to build and simulate the models rather than dealing with the technicality of formalisms used to build and modify the models.

**C-IMMSIM, a generic immune system simulation platform** (F. Castiglione)
The computer model C-IMMSIM can be seen as the outcome of a collaborative effort between a biologist, who provides insights into mechanisms and actions, and a mathematician, who translates that knowledge into a quantitative framework [47]. Developing an accurate computer model that represents the complexity of the immune system and produces meaningful outcomes is a challenging task. However, by accepting necessary approximations and building upon solid theoretical mathematical and biological assumptions, along with personalized data to infer the model parameters [28], the C-IMMSIM model can be considered as an underlying generic model of an individual's immune digital twin.

The essential components and prerequisites that have influenced the development of C-IMMSIM are: diversity in specific repertoires; probabilistic actions capturing the inherent stochasticity of many mechanisms; cooperation between different cell types; cell movement and global control; specific cell-cell and cell-molecule interactions; competition and memory cells; clonal selection and proliferation; controls and memory.

All these elements have been incorporated into the C-IMMSIM model using specific mathematical or algorithmic choices. The model can be categorized as an Agent-Based Model (ABM), where individual cells are represented with their unique attributes, such as position, age, membrane receptors, activation status, or differentiation state. ABMs are well-suited for simulating the immune system due to their ability to handle stochastic actions, cell movement, and individual dynamics, while allowing large populations to be simulated and tracked. During the simulation, cells undergo transitions between activation or differentiation states, influenced by stochastic events that rely on the compatibility of their binding sites. While simulating billions of agents and incorporating anatomical variations and an individual's immunological history remains impractical, even with high-performance computers, the overall state of the system in the simulation can still be considered a representative immunological state for an individual. In essence, by adopting a digital twin perspective, the model can be tailored to match a patient's physical attributes and current health condition. Consequently, it can offer valuable insights into an individual's immune status and potential outcomes when encountering specific stimuli.

**Toward a medical digital twin for pneumonia patients in the Intensive Care Unit** (R. Laubenbacher, B. Mehrad)
Doctors in intensive care units (ICUs) make decisions in a complex environment, bombarded with thousands of pieces of data, and often under intense time pressure and heterogeneity of patient response to treatment. Available ICU risk calculators provide highly accurate predictions of a patient's length of stay and likelihood of death, but do not provide actionable information about what interventions could be applied to an individual patient to improve the outcome. A common condition of ICU patients is pneumonia. It is the second most common cause of hospital admissions (after admissions for childbirth), with up to 10% of patients requiring an ICU stay. And up to 5% of hospital patients contract pneumonia. It is the leading cause of death worldwide for children under 5. The goal of this project is to build a pneumonia digital twin for ICU patients that serves as a decision support tool for the doctor. The aim of this project is to construct a personalized computational model that encodes disease-relevant biological mechanisms and is dynamically recalibrated as new patient data become available. The computational model underlying the pneumonia MDT will be an extension and modification of a model of the early immune response to a respiratory fungal infection, using the fungus *Aspergillus fumigatus* as the model pathogen [48].

Ongoing work includes a study of the early immune response to viral and bacterial pathogens. These studies will be used to expand the computational model for fungal pneumonia, covering all major pathogens causing pneumonia. A tissue culture platform, combined with a cryopreservation technique that keeps human lung tissue functional over several days is being used with lung tissue obtained from surgeries. Collecting heterogenous data from infected tissue from a range of donors allows us to "personalize" the computational model to different donors and investigate heterogeneity in disease progression and response to drugs. This represents the next step in developing the computational model to a state where it can be personalized to actual patients. A part of future work to be done is to integrate this tissue/organ-scale model with a physiological model that allows implementation of all standard treatments available to a pneumonia patient, allowing the comprehensive simulation of patient trajectories under treatment. The final product will be an MDT that is based on a mechanistic computational model, is calibrated dynamically to a pneumonia patient in the ICU and can be used to optimize the patient's treatment.

**A breast cancer digital twin** (T. Yankeelov)
Yankeelov and colleagues have developed mechanism-based mathematical models that are initialized and calibrated with patient-specific, quantitative imaging data for a variety of cancers, especially the breast. The imaging data has included both quantitative positron emission tomography [49] and magnetic resonance imaging (MRI) [50], with a particular emphasis on dynamic contrast enhanced MRI to report on blood flow, and diffusion weighted MRI to report on cellularity. (Using medical imaging data has the advantage of being able to report on anatomical, physiological, cellular, and molecular data non-invasively and at multiple time points to update a digital twin throughout the course of therapy [51]. Given a high-resolution anatomical image to establish the computational domain, reaction-diffusion equations accounting for tissue mechanical properties and therapeutic regimens are solved over the breast to establish patient specific parameters related to tumor cell migration, tumor proliferation, and response to therapy. Once the model system is calibrated, it can be run forward in time to predict the spatio-temporal response of the tumor to the specific treatment with high accuracy [52]. Given that the model can faithfully predict the spatial and temporal dynamics of an individual tumor, it is natural to use it to form the backbone of an MDT designed to predict and, ultimately, identify therapeutic regimens to optimize the tumor response. In fact, preliminary simulation results indicate that merely delivering the same total dose in a patient-specific way can potentially improve outcomes [53]. With the recent inclusion of pembrolizumab into the standard-of-care for the treatment of triple negative breast cancer, the ability to simulate which patients would benefit from immunotherapy and which patients should avoid it (and the associated side effects) is difficult to overstate.

It is important to note that only by employing mechanism-based models can one simulate a range of therapeutic options, including new emerging therapeutics without large clinical trials to use as training data. When using a strictly data-driven approach, one can only search for responses to therapeutic regimens that are included in the training set. By using a mechanism-based model, one is not limited to only the therapeutic regimens included in a historical training set.

**A leukemia digital twin** (I. Shmulevich)
The Acute Myeloid Leukemia Digital Twin (AML-DT) project is an initiative funded by the National Cancer Institute (NCI) and the Academy of Finland. It aims to develop a comprehensive digital twin system for AML. This project is characterized by its unique approach that combines disease-specific knowledge graphs instantiated with patient data, machine learning, and mechanistic models. These include gene regulatory network models and multicellular models of hematopoiesis and leukemogenesis, which are designed to incorporate key mechanisms of cancer progression. The overarching goal of this project is to predict disease progression and optimize response to therapies, thereby revolutionizing the way we understand and treat AML. The project is a collaborative effort, bringing together diverse fields such as modeling, machine learning, human-computer interaction, and clinical practice.

The development of the AML digital twin necessitates a variety of patient data, including clinical data, flow cytometry measurements, cytogenetics, and mutation panels. These data are utilized to individualize each digital twin, creating a personalized representation of the patient's disease state. Alongside patient-specific data, the project also incorporates public datasets for the construction of knowledge graphs. These datasets include *ex vivo* drug sensitivity data and molecular profiling, both linked with clinical outcomes. The integration of individual patient data and public datasets enhances the digital twin's ability to predict disease progression and drug response, which is the primary objective of this project.

The key aspects of the immune system relevant for AML are captured by the digital twin through the integration of detailed domain-specific knowledge graphs with multiscale dynamical models of the tumor microenvironment. These models incorporate key mechanisms of cancer progression, which can aid in the development of new therapies.

The digital twin approach goes beyond being just a model. Each AML patient will have a digital twin individually tailored using information produced in a clinical laboratory. This is combined with a model-based approach for making personalized predictions. An important aspect of this approach is the learning-cycle, where patient outcomes are continuously utilized to improve predictions. Over time, the system will improve as discoveries are made related to the biological aspects that are most important for accurate prediction of patient outcomes. This approach allows for a dynamic and evolving representation of the patient's disease state, providing a more accurate and personalized prediction of disease progression and treatment response.

| Presenter | Modeling framework | Disease | Data Structure of the real world object | Data Link b/w in silico & real world | Specification | "Personalization" | Purpose | Patient facing interface |
|---|---|---|---|---|---|---|---|---|
| Yankeelov | Custom | Cancer | Multimodal - Molecular, Imaging and clinical data | Yes | Mechanistic Eq PDE/ODE | Yes, from data collected from an intervention | Prognosis, Optimization of existing therapy, control discovery | Maybe |
| Shmulevich | | AML | Multimodal - Molecular & clinical/epidemiological | Yes | Mechanistic model w/ ML component for feature aggregation | Initialization from a virtual population, collected data from the patient over time | Prognosis, optimize therapy, control discovery | Yes |
| Kirschner | Custom | TB | Molecular, imaging & clinical (animal & patient) | Yes | Mechanistic Model (Multiscale ABM w/ODEs) | Explored possibility space of outcomes & created | Prognosis, optimization of existing therapies, control | No, evolving |

| | | | | | | virtual population | discovery of vaccine testing | |
|---|---|---|---|---|---|---|---|---|
| Smith | Custom | Immune system (Viral infection) | Molecular & clinical | Yes | Mechanistic Model | Virtual populations | Prognosis, scientific discovery | No |
| Helikar | Custom | Immune system | Molecular & clinical | Evolving | Mechanistic Model | Evolving, can create a VP | Prognosis, Optimization | Planned |
| Ziemssen | Custom | MS | Molecular, imaging & clinical | Yes | Evolving | Data structure map | Classification, prognosis & optimization | Yes |
| Castiglione | C-IMMSIM | Immune system | Molecular & clinical | Evolving | Mechanistic Model | Evolving | Prognosis, scientific discovery, control discovery and in silico trials | No |
| Macklin | Physicell | General purpose | Molecular, cellular, and tissue | Evolving. Recent integration with spatial transcriptomics | Mechanistic Model | Evolving. Agent parameters can be calibrated to individual patient characteristics | Prognosis, optimization & control discovery | No |
| Glazier | CompuCell3D | | Molecular | Evolving | Mechanistic Model | Evolving | Prognosis, optimization & control discovery | No |
| Jett-Tilton | | PTSD | Genomic & clinical | Yes | No, evolving | Yes | Prognosis, optimization & clinical trials | No, evolving |
| Laubenbacher, Mehrad | Custom | Aspergillosis | Molecular & clinical | Yes | Mechanistic Model (Multiscale ABM w/Boolean networks) | Yes, w/ MLed parameters | Prognosis, optimize therapy, control discovery | No. Doctor facing interface |

**Table 1.** MDT projects in different stages of progress by Forum participants. The columns contain information about the technical features of the MDT, the type of data required for calibration, and the application specifications.

**CONCLUSION**

The data from an individual patient captures different aspects of their characteristics and health status. We have genomic data, gene expression measurements, protein, and metabolite concentrations in different tissues under different conditions, imaging data of everything from immune cells in lymph nodes to functional MRI data in the brain, electronic health records, to lifestyle and behavioral data. They all provide information about some aspect of a person, and the challenge is to integrate them in a meaningful way to provide a holistic representation. A computational model of the patient that is dynamically updated with all this information is a natural, and maybe the only, way to accomplish the data integration required. The confluence of several simultaneous developments has created an environment in which this promise of personalized medicine is taking on shape: vastly increased availability of data, from the molecular to the population scale, leading to a deeper understanding of human biology and its role in health and disease, and, finally, an expansion of our computational and modeling tools.

      A well-designed funding program for MDT research by the public sector is crucial if substantial progress is to be made over the next decade. Public sector agencies like the National Institutes of Health and the National Science Foundation, as well as the U.S. Department of Defense, can play a crucial role in creating an interdisciplinary research ecosystem that brings together the needed expertise. New funding paradigms should be considered for this purpose. For instance, the structure of the Horizon research programs funded by the European Union is well-suited to MDT projects that rely on the development of a collection of parts that together assemble to an MDT, but do not necessarily have a research rationale of their own. There can also be an important role for the business community and philanthropic organizations in providing funding for this effort and collaborating on the myriad research problems that will need to be tackled and solved. The Forum we are reporting on here is intended to support a dialog around this topic. Collectively, a community is emerging around this effort that can, with the right resources, help make rapid progress on bringing MDTs to patients at a large scale.

**ACKNOWLEDGEMENTS**
RL: U.S. Army ACC- APG- RTP W911NF, NIH 1 R01 HL169974-01, U.S. DoD DARPA HR00112220038, NIH 1 R011AI135128-01, NIH 1 R01 HL169974-01
FA: None
GA: NIH UO1EB025825, Department of Defense CDMRP- W81XWH-22-MBRP-CTRA, DARPA HR00111950027
FC: None
SE: None
LF: U.S. DoD DARPA HR00112220038
TH: NIH Grant #R35GM119770, University of Nebraska-Lincoln Grand Challenges Catalyst Award
MJT: None
DK: NIH R01 AI50684
PM: NSF 1720625, Jayne Koskinas Ted Giovanis Foundation for Health and Policy, Leidos Biomedical Research contract no. 75N91019D00024
BMe: NIH 1 R01 HL169974-01, NIH 1 R011AI135128-01, NIH 1 R01 HL169974-01
BMo: NIH R35HL144481, NIH DK124290
VP: None
IS: NIH NCI R01CA270210
AS: NIH R01 AI170115, NIH R01 AI139088
TY: NCI 1U01CA253540, NCI 1R01CA240589
TZ: None

The authors are grateful to Ms. Lisa Oppel who was in charge of the Forum organization and all logistics.

**COMPETING INTERESTS**
None: RL, FA, GA, FC, SE, LF, JG, MJT, DK, PM, BMe, BMo, VP, IS, AS, TY, TZ
TH: majority stakeholder in ImmuNovus, Inc and Discovery Collective, Inc.

**AUTHOR CONTRIBUTIONS**
RL secured funding and organized the Forum, coordinated, and contributed to writing the manuscript. All other authors contributed to the scientific content of the meeting and participated in preparing the manuscript.